\documentclass[twocolumn]{aastex63}

\usepackage{graphicx} % Required for including pictures
\usepackage{amsmath}
\usepackage{ amssymb }

\usepackage[utf8]{inputenc}
\usepackage[T1]{fontenc}

\usepackage{color}

\usepackage{float} % Allows putting an [H] in \begin{figure} to specify the exact location of the figure
\usepackage{wrapfig} % Allows in-line images such as the example fish picture

\usepackage[makeroom]{cancel} %allows strikethroughs
\usepackage{comment}

\usepackage{lipsum} % Used for inserting dummy 'Lorem ipsum' text into the template

\usepackage{outlines}
\usepackage{amsmath}

\graphicspath{{./figs/}} % Specifies the directory where pictures are stored

\newcommand{\beq}{\begin{equation}}
\newcommand{\eeq}{\end{equation}}

\newcommand{\bvec}{\begin{pmatrix}}
\newcommand{\evec}{\end{pmatrix}}
\newcommand{\lp}{\left(}
\newcommand{\rp}{\right)}
\newcommand{\pa}[2]{\frac{\partial #1}{\partial #2}}

\newcommand{\paf}[2]{\partial #1 / \partial #2}

\newcommand{\ve}[1]{\mathbf{#1}}

\newcommand{\Gammav}{\boldsymbol{\Gamma}}

\begin{document} %apj

%----------------------------------------------------------------------------------------
%	TITLE PAGE
%----------------------------------------------------------------------------------------

\title{Synchrotron-driven instabilities in relativistic plasmas of arbitrary opacity}

% apj
\author[0000-0002-6002-9169]{Ian E. Ochs}
\email{iochs@princeton.edu}
\affiliation{Department of Astrophysical Sciences, Princeton University, Princeton, New Jersey 08540, USA}

%jpp 

%\shorttitle{Synchrotron-driven instabilities in relativistic plasmas of arbitrary opacity}
%\shortauthor{I.E. Ochs}

%\author{Ian E. Ochs\aff{1}
%	\corresp{\email{iochs@princeton.edu}}}
%\affiliation{\aff{1}Department of Astrophysical Sciences, Princeton University, Princeton, New Jersey 08540, USA}

%\date{\today}% It is always \today, today,
%  but any date may be explicitly specified

%\begin{document} %jpp
%\maketitle %jpp

\begin{abstract}
		Recent work has shown that synchrotron emission from relativistic plasmas leads the electron distribution to form an anisotropic ring in momentum space, which can be unstable to both kinetic and hydrodynamic instabilities.
		Fundamental to these works was the assumption that the plasma was optically thin, allowing all emitted radiation to escape.
		Here, we examine the behavior of these instabilities as the plasma becomes more optically thick.
		To do this, we extend a recently-developed Fokker-Planck operator for synchrotron emission and absorption in mildly relativistic plasmas to fully relativistic plasmas.
		For a given set of plasma parameters, photons emitted by higher-energy electrons tend to be higher frequency, and thus more easily escape the plasma. 
		As a result, the ratio of the photon emission rate (radiative drag) to absorption rate (radiative diffusion) for a given electron is extremely energy-dependent.
		Given this behavior, we determine the critical parameters that control the opacity, and show how the plasma gradually transitions to become more isotropic and stable at higher opacity.
\end{abstract}

%\vspace{3pt}

%\tableofcontents

\section{Introduction}

Recently, several papers \citep{Bilbao2023RadiationReaction,Bilbao2024RingMomentum,Zhdankin2023SynchrotronFirehose} have demonstrated the importance of synchrotron radiation in driving plasma instabilities.
The basic mechanism is that the radiative drag due to synchrotron emission, which depends strongly on the perpendicular momentum $u_\perp \equiv p_\perp / mc^2$, drives significant anisotropy in the electron distribution.
This produces a ``ring distribution,'' which is both kinetically unstable due to population inversion along $u_\perp$ with $\paf{f}{u_\perp} > 0$ \citep{Bilbao2023RadiationReaction,Bilbao2024RingMomentum}, and can also be hydrodynamically unstable due to pressure tensor anisotropy with $P_\parallel > P_\perp$ \citep{Zhdankin2023SynchrotronFirehose}.

However, these studies have typically assumed an optically thin plasma, where photons are emitted but never reabsorbed.
This is undoubtedly the case for small, tenuous plasma systems; but as plasma systems become larger and denser, the opacity increases, and radiation reabsorption can become significant.
In this paper, we aim to determine the fundamental parameters that determine the plasma opacity, and to model how the tendency toward instability is modified by increasing plasma opacity.

Emission-absorption modeling of synchrotron radiation is complicated by the fact that the emission occurs at a range of frequencies, and each of these frequencies is absorbed at a different rate in the plasma.
Thus, the plasma tends to remain opaque at low frequencies even as it becomes transparent at high frequencies.
Recently, a Fokker-Planck operator was developed to account for this frequency-dependent plasma opacity \citep{Ochs2024ElectronTail}, due to the importance of synchrotron power losses in aneutronic fusion \citep{Mlodik2023SensitivitySynchrotron}.
However, the operator relies on a sum over harmonics which becomes onerous as the plasma becomes substantially relativistic.
Here, we take advantage of the near-continuous spectrum of synchrotron emission in the highly relativistic limit to extend the operator to the relativistic region of interest for astrophysics and high-energy-density experiments, which was the focus of the recent work on synchrotron-driven instabilities.

The paper is thus laid out as follows.
In Section~\ref{sec:Model}, we review the mixed-opacity Fokker-Planck model, adapting it to the relativistic limit and showing how it depends on a single opacity function $\lambda(\ve{u})$.
In Section~\ref{sec:OpacityFunctions}, we show how to calculate the opacity functions for a relativistic plasma, completing the model.
In Section~\ref{sec:Heuristics}, we describe the approximate behavior of the diffusion, with an emphasis on the timescales and the quasi-steady-state towards which the system evolves, showing that some degree of ring distribution and pressure anisotropy persist even in fairly opaque plasma conditions.
Finally, in Section~\ref{sec:Simulations}, we perform numerical simulations, showing how an initial Maxwell-J\"uttner distribution evolves for plasmas of different opacities.

%\begin{itemize}
%	\item Recent papers have shown that synchrotron radiation drives plasmas unstable, either via phase space instability in $p_\perp$ \cite{Bilbao2023RadiationReaction} or pressure anisotropy $p_\parallel / p_\perp > 1$ \cite{Zhdankin2023SynchrotronFirehose}.
%	\item These studies assumed an optically thin plasmas.
%	\item Important to understand when this assumption is justified, and what the behavior looks like when plasma grows optically thicker.
%	\item Recently developed reduced model for synchrotron emission and absorption in mixed opacity plasma.
%	\item Extend this model to ultra-relativistic regime.
%\end{itemize}

\section{Synchrotron Diffusion Operator For Mixed-Opacity Relativistic Plasma} \label{sec:Model}

As in \cite{Ochs2024ElectronTail}, we model a uniform plasma under the emission and absorption of synchrotron radiation.
The fundamental assumption of the model is that beneath a certain cutoff frequency $\omega^*$, the plasma is optically thick, and thus the photons are in thermal equilibrium with the plasma, while photons emitted at frequencies above $\omega^*$ escape the plasma.
Thus, the photon spectrum is blackbody below $\omega^*$, and 0 above $\omega^*$.
As a result, emission and absorption of low-frequency photons lead to thermally-balanced drag and diffusion, while emission of high frequency photons leads to unbalanced radiation drag.

Under the assumptions above, the impact of synchrotron radiation on electrons in a plasma with a constant magnetic field can be modeled by a Fokker-Planck operator in coordinates $\ve{u}\equiv (u,\psi)$, where $u = p/mc$ is the relativistically-normalized momentum, and $\psi = \arccos (u_\parallel / u)$ is the electron pitch angle.
The operator takes the form:
\begin{align}
	\pa{f}{t} &= \frac{1}{\sqrt{g}}\pa{}{\ve{x}} \cdot \left[ \sqrt{g} \left(- \Gammav f + \ve{D} \cdot \pa{f}{\ve{x}} \right) \right] \label{eq:FundamentalDiffusion},
\end{align}
where
\begin{align}
	\Gammav &= -\nu_{R0} \frac{\gamma_\perp^2}{\gamma} \ve{g}; \label{eq:GammaUALpha}\\
	\ve{D} &= (1-\lambda(\ve{u})) \nu_{R0} \chi_\text{bb} \left( \frac{\gamma_\perp^2}{\gamma} \right)^2 \frac{1}{u_\perp^2} \left(\ve{g}\ve{g} + \Delta \ve{h}\ve{h} \right); \label{eq:DUAlpha}
	\\
	\nu_{R0} &= \frac{2}{3} \frac{e^2 \Omega_0^2}{m c^3}; \quad \sqrt{g} = x^2 \sin \psi. \label{eq:RadiationFrequencyAndMetric}
\end{align}
Here, $\chi_\text{bb} = T_\text{bb} / mc^2$ is the normalized blackbody temperature of the synchrotron radiation (usually taken to be the average electron temperature), and $\gamma \equiv \sqrt{1+u^2}$ and $\gamma_\perp = \sqrt{1+u_\perp^2}$ are the Lorentz factors associated with the total and perpendicular momenta respectively.
The vectors which constitute the diffusion tensor outer products are given by:
\begin{align}
	\ve{g} = \gamma_\perp^{-2} \bvec \gamma^2 u \sin^2 \psi \\ \cos\psi \sin \psi \evec ; \quad \ve{h} = \bvec 0 \\ 1 \evec.
\end{align}
The effects of frequency-dependent opacity are contained in the opacity function $\lambda(\ve{u})$, which represents the fraction of synchrotron radiation, emitted by particles at momentum $\ve{u}$, that escapes the plasma.

In the ultrarelativistic limit ($u \gg 1$) considered here, many simplifications occur.
The pitch angle scattering coefficient $\Delta \rightarrow 0$, and furthermore $\gamma \rightarrow u$, $\gamma_\perp \rightarrow u_\perp$.
This makes $\ve{g} \rightarrow \ve{u}$, resulting in:
\begin{align}
	\Gammav &= -\nu_{R0} u_\perp^2 \hat{u} \label{eq:GammaUltraRelativistic}\\
	\ve{D} &= (1-\lambda) \nu_{R0} \chi_\text{bb} u_\perp^2 \hat{u}\hat{u} . \label{eq:DUltraRelativistic}
\end{align}

Eqs.~(\ref{eq:GammaUltraRelativistic}-\ref{eq:DUltraRelativistic}) show that the net effect of the synchrotron radiation is one-dimensional diffusion and drag along $\hat{u}$.
Each of \cite{Bilbao2023RadiationReaction,Bilbao2024RingMomentum,Zhdankin2023SynchrotronFirehose} considered the limit of $\lambda \rightarrow 1$, where all radiation escapes, and the Fokker-Planck equation becomes an advection equation for the radiative drag.
Conversely, in the opposite limit $\lambda \rightarrow 0$, i.e. a blackbody radiation spectrum, the steady-state solution Eq.~(\ref{eq:FundamentalDiffusion}) is the relativistic limit of the Maxwell-J\"uttner distribution:
\begin{align}
	f_e \propto e^{-u/\chi_\text{bb}}.
\end{align}
However, note that the lack of pitch-angle scattering means that each value of $\psi$ will in general have a different constant of proportionality.
This result is analogous to other problems of constrained entropy maximization, where the distribution can relax to a Gibbs distribution only while respecting invariants of the dynamics \citep{Kolmes2020MaxEntropy}.

Many plasmas, however, will be in neither the optically-thick nor optically-thin limit, but in an intermediate regime of $0 < \lambda(\ve{u}) < 1$.
To complete the model, it is necessary to calculate the opacity function $\lambda(\ve{u})$ for a relativistic distribution of electrons.
For relativistic plasmas, this calculation requires a fundamentally different approach than for mildly relativistic plasmas.
%Because the harmonic-sum approach of Ref.~\todo{other} becomes onerous for relativistic plasmas, the calculation of $\lambda$ requires fundamentally different techniques in the relativistic limit.

\section{Opacity Function in an Ultrarelativistic Plasma} \label{sec:OpacityFunctions}

In order to make use of the synchrotron Fokker-Planck model, we must solve for the opacity function $\lambda$ in an ultrarelativistic plasma.
The approach taken in \cite{Ochs2024ElectronTail} was to sum emission over the different cyclotron harmonics and calculate $\lambda$ based on the fraction of emission above a cutoff harmonic.
However, this approach quickly becomes untenable in the ultrarelativistic limit, as even at $u = 5$, 1000 harmonics are needed, and the number of harmonics required scales roughly as $u^3$ \citep{Melrose1980PlasmaAstrophysics}.
Luckily, new approximations are available in the highly relativistic limit that make the calculation much more analytically tractable.

\subsection{Ultrarelativistic $\lambda$}

In general, the opacity of a plasma to synchrotron radiation is a strong function of the frequency.
Thus, radiation below some cutoff frequency $\omega^*$ is trapped, while radiation above $\omega^*$ escapes, so that to a good approximation the fraction of radiation emitted by an electron that escapes the plasma is given by:
\begin{align}
	\lambda (\ve{u}) \equiv \frac{\sum_{i,s} 2\pi  d\omega \int_{-1}^{1} d \cos \theta \int_{\omega^*}^\infty \eta_s^{(i)} (\ve{u})}{\sum_{i,s} 2\pi  \int_{-1}^{1} d \cos \theta \int_0^\infty d\omega \eta_s^{(i)}(\ve{u})}. \label{eq:LambdaGeneral}
\end{align}
Here, $i$ is the plasma mode label, $s$ is the harmonic, and $\eta_s^{(i)}(\ve{u})$ is the emissivity at the $s^\text{th}$ harmonic of mode $i$ by an electron with momentum $\ve{u}$.

%In general, $\omega^*$ is a function of $\theta$, the angle of the emitted photon.
%However, in the relativistic limit, the headlight effect ensures that the radiation angle $\theta$ is approximately equal to the pitch angle $\psi \equiv \arccos (u_\parallel / u)$, with the spread in the angle of emission of order $\Delta \theta = 1/\gamma$ (\cite{Bekefi1966RadiationProcesses} pg. 191).

%Ref.~\cite{Bekefi1966RadiationProcesses} (6.34), (6.35), (6.38), (6.4)
%and Ref.~\cite{Melrose1980PlasmaAstrophysics}.
Radiation is emitted in two modes, the O and X modes.
It is known that the total power density radiated per unit frequency, summing over both modes, from a relativistic electron is given by \citep{Bekefi1966RadiationProcesses,Melrose1980PlasmaAstrophysics}:
%\begin{align}
%	2 \pi \sum_s \int_{-1}^{1} d \cos \theta \, \eta_s^{(i)}(\ve{u}) &= \frac{\sqrt{3} m_e \omega_{p0}^2 \bar{\omega}_b}{8 \pi^2 c} \frac{\omega}{\omega_c} \int_{\omega/\omega_c}^{\infty} K_{5/3}(t) dt, \label{eq:EmissionPerUnitFrequency}
%\end{align}
\begin{align}
	2 \pi \sum_s \int_{-1}^{1} d \cos \theta \, \eta_s^{(i)}(\ve{u}) &= \frac{\sqrt{3} e^2 \bar{\omega}_b}{2 \pi c} \frac{\omega}{\omega_c} \int_{\omega/\omega_c}^{\infty} K_{5/3}(t) dt, \label{eq:EmissionPerUnitFrequency}
\end{align}
where:
\begin{align}
	\omega_c &= \frac{3}{2} \bar{\omega}_b \gamma^2 \label{eq:omegaC}\\
	\bar{\omega}_b &= \Omega_0 (1 - \beta_\parallel^2)^{1/2} \approx \Omega_0 \sin \psi, \label{eq:omegaBBar}
\end{align}
and where the last equality comes from $\beta_\parallel \approx u_\parallel / u = \cos\psi$ in the ultrarelativistic limit.

If this radiation came out at a broad spread of $\theta$, this result would not be useful, since to calculate $\lambda$ we need to integrate $\eta_s^{(i)}$ over $\omega$ with a lower limit $\omega^*$ that depends on $\theta$.
However, since all radiation is emitted at approximately the pitch angle $\psi$, there is only one $\theta$ (and thus one $\omega^*$) for any $\ve{u}$, i.e. $\omega^*(\theta)$ is replaced with $\omega^*(\psi)$.
This allows us to switch the order of integration in Eq.~(\ref{eq:LambdaGeneral}) and plug in the pre-integrated result from Eq.~(\ref{eq:EmissionPerUnitFrequency}).
We then have:
\begin{align}
	\lambda(x^*) &= \frac{\int_{x^*}^\infty dx \bar{\eta}(x)}{\int_{0}^\infty dx \bar{\eta}(x)}, \label{eq:LambdaUltraRelativistic}
\end{align}
where
\begin{align}
	\bar{\eta} (x) &= x \int_x^\infty dt K_{5/3}(t) dt,
\end{align}
and $x^* = \omega^* / \omega_c$.
Thus, we see that $\lambda(\ve{u})$ depends on the single parameter $x^*(u,\psi)$, which represents an appropriately normalized frequency cutoff.

\subsubsection{Asymptotically-Valid Fit for $\lambda$}

It is useful to derive an approximate analytic form for $\lambda$.
To start, we examine the asymptotic limits of $\lambda$ at low and high $x^*$.
The first limit is easy: as $x^* \rightarrow 0$, $\lambda \rightarrow 1$.

To see the behavior of $\lambda$ in the high-$x^*$ limit, we first note that the denominator of Eq.~(\ref{eq:LambdaUltraRelativistic}) is given by
\begin{align}
	\int_{0}^\infty dx \bar{\eta} (x) = 1.61227.
\end{align}
To find the numerator, we start from Ref.~\cite{Bekefi1966RadiationProcesses}'s Eq.~(6.37):
\begin{align}
	\bar{\eta} (x) \approx \sqrt{\frac{\pi}{2}} x^{1/2} e^{-x} \text{ for } x \gg 1. \label{eq:K53IntegralApproximationHighX}
\end{align}
As a result, the numerator of Eq.~(\ref{eq:LambdaUltraRelativistic}) is approximately given by:
\begin{align}
	\int_{x^*}^\infty dx \bar{\eta} (x) =  e^{-x^*} \left((x^*)^{1/2} + \mathcal{O}((x^*)^{-1/2})\right)
\end{align}

\begin{figure*}
	\centering
	\includegraphics[width=\linewidth]{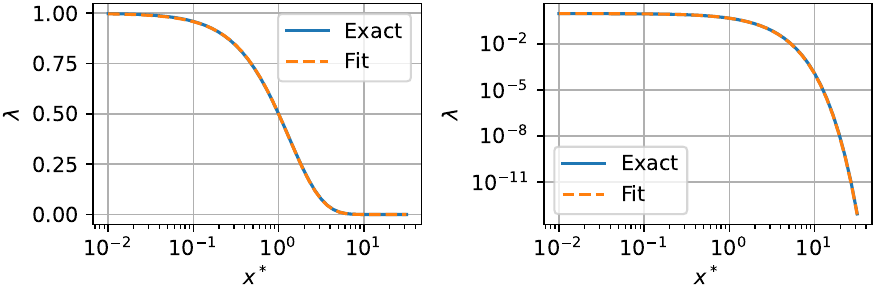}
	\caption{Escaping radiation fraction $\lambda$ as a function of normalized lowest escaping frequency $x^*$: exact result (Eq.~\ref{eq:LambdaUltraRelativistic}) and fit (Eq.~\ref{eq:UltrarelativisticLambdaFit}).
		The right plot shows the same result on a log-log plot, demonstrating asymptotic agreement for large $x^*$ (small $\lambda$).}
	\label{fig:UltrarelativisticLambda}.
\end{figure*}

To agree in both asymptotic limits, we can fit a function of the form:
\begin{align}
	\lambda_\text{fit} \equiv e^{-x^*} \left(\frac{1 + (x^*)^2}{1 + a (x^*)^{b} + 1.612 (x^*)^{3/2}}  \right). \label{eq:UltrarelativisticLambdaFit}
\end{align}
This produces a very accurate fit for $a = -1.140$, $b = 1.070$ (Fig.~\ref{fig:UltrarelativisticLambda}).
Note that it is more important for $\lambda$ to fit at $\lambda \ll 1$, since this determines the delicate balance of drag and diffusion that leads to accurate loss rates in the optically thick limit, than it is to fit $(\lambda - 1)$ at $\lambda \rightarrow 1$, where the plasma is optically thin and most of the energy is lost on a very short timescale.

\subsection{Cutoff frequency}

Having found $\lambda(\ve{u})$ to be a function of the normalized cutoff frequency $x^* = \omega^*/\omega_c$, it is now important to be able to calculate $\omega^*$ as a function of the plasma parameters.
Usually, the cutoff $\omega^*$ for a mode $i$ is found by setting 
\begin{align}
	\alpha^{(i)}(\omega^*,\theta) L = (\sin \theta)^a,
\end{align} 
where $a \approx 1$ for a slab or cylinder of plasma and $a \approx 0$ for a sphere. 
Here, $\alpha^{(i)}(\omega,\theta)$ is the absorption coefficient, given in the classical ($\hbar \omega \ll 1$) limit for a distribution function that depends only on normalized energy $\gamma$ via Kirchoff's law by (see e.g. \cite{Bekefi1966RadiationProcesses} Eq.~(2.46)):
\begin{align}
	\alpha^{(i)}(\omega,\theta) &= - \frac{8 \pi^3 n_e}{m_e \omega^2} \int d^3\ve{u} \pa{f_e}{\gamma} \eta^{(i)}(\omega,\theta,\ve{u}),
\end{align}
where $f_e$ is the electron distribution function in normalized momentum space $\ve{u}$.
Since in the highly relativistic limit $u \approx \gamma$, and since the distribution has spherical symmetry, we can write (noting that $d^3\ve{u} = 2\pi u^2 d\cos \psi du$):
\begin{align}
	\alpha^{(i)}(\omega,\theta) &= - \frac{16 \pi^4 n_e}{m_e \omega^2} \int d u \, d \cos \psi \, \pa{f}{u} \eta^{(i)}(\omega,\theta,\ve{u}).
\end{align}

To derive a clean expression for $\alpha$, we will make use of the headlight approximation.
We will also neglect the difference between the O and X modes, treating them as a single mode with the combined emission characteristics from Eq.~(\ref{eq:EmissionPerUnitFrequency}).
This approximation becomes increasingly accurate at high $x = \omega/\omega_c$, where the emission is almost all in the X mode, and never misestimates $\alpha$ by more than a factor of 2, which (as we will see later) does not have a significant impact on the cutoff frequency.
Thus, we will drop the mode superscript, and write:
\begin{align}
	\eta(\omega,\theta,\ve{u}) = W(\omega,u,\psi) \delta(\cos \theta - \cos \psi), \label{eq:HeadlightEta}
\end{align}
yielding
\begin{align}
	\alpha(\omega,\theta) &= -\frac{16 \pi^4 n_e}{m_e \omega^2} \int_0^\infty du \, u^2 \pa{f}{u} W(\omega,u,\psi=\theta).
\end{align}
Here, $W(\omega,u,\psi)$ is consistently solved for by plugging the form of  Eq.~(\ref{eq:HeadlightEta}) into Eq.~(\ref{eq:EmissionPerUnitFrequency}), yielding:
%\begin{align}
%	W(\omega,\psi) &= \frac{\sqrt{3} m_e \omega_{p0}^2 \bar{\omega}_b}{16 \pi^3 c n_e} \bar{\eta} (x). \label{eq:WFunction}
%\end{align}
\begin{align}
	W(\omega,u,\psi) &= \frac{\sqrt{3} e^2 \bar{\omega}_b}{4 \pi^2 c} \bar{\eta} (x). \label{eq:WFunction}
\end{align}
%It looks like $\psi$ does not enter here, but in fact it does, through $x = \omega / \omega_c$, which depends on $\bar{\omega}_b = \Omega_0 \sin \psi$ [Eqs.~(\ref{eq:omegaC}-\ref{eq:omegaBBar})].
Thus, for a general distribution, we have:
\begin{align}
	\alpha(\omega,\theta) &= -\frac{\sqrt{3}\pi \omega_{p0}^2 \Omega_0 \sin \theta}{ \omega^2 c } \int_0^\infty du \, u^2 \pa{f}{u}  \bar{\eta} (x);\\
	x &\equiv \frac{2\omega}{3u^2 \Omega_0 \sin \theta},
\end{align}
where $\omega_{p0} = \sqrt{4 \pi n_e e^2 / m_e}$ is the electron plasma frequency.  

To proceed further, we will assume that the bulk electron distribution is a Maxwell-J\"uttner distribution with a normalized temperature $\chi_e \equiv T_e / mc^2$:
\begin{align}
	f_e \equiv \frac{1}{4\pi \chi K_2 (\chi^{-1})} e^{-\gamma/\chi_e} \approx \frac{1}{8\pi \chi_e^3} e^{-u/\chi_e}.
\end{align}
Thus, $\alpha$ takes the simple form:
\begin{align}
	\alpha(\omega,\theta) &= \frac{\sqrt{3}\omega_{p0}^2 \Omega_0 \sin \theta}{ 8 \omega^2 c \chi_e^4} \int_0^\infty du \, u^2 e^{-u/\chi_e}  \bar{\eta} (x)\\
%	&= \frac{\sqrt{3}\omega_{p0}^2 \Omega_0 \sin \theta}{ 8 \omega^2 c \chi_e} I_1,
	&= \frac{1}{6\sqrt{3}}\frac{\omega_{p0} }{\Omega_0} \frac{\omega_p}{c} \frac{1}{\chi_e^5 \sin \theta} \frac{I_1}{y^2}, \label{eq:alphaFromY}
\end{align}
where
\begin{align}
	I_1 &= \int_0^\infty dw \, w^2 e^{-w}  \bar{\eta} (x) \label{eq:I1Exact}\\
	x &\equiv y/w^2 \label{eq:xFromy} \\
	y &\equiv \frac{2\omega}{3\chi_e^2 \Omega_0 \sin \theta},
\end{align}
and $w \equiv u / \chi_e$.
Thus, we see that $I_1$ is a function only of $y$.

We can now easily find the normalized cutoff frequency $x^*$.
Examining Eq.~(\ref{eq:xFromy}), we see that $x^* = y^*/w^2$ is a simple function of $y^*$, where $y^*$ is the solution to $\alpha L = \sin^a \theta$, using $\alpha$ as defined in Eq.~(\ref{eq:alphaFromY}).
Since the headlight approximation equates the radiation angle $\theta$ with the pitch angle $\psi$, the equation for $y^*$ becomes:
\begin{align}
	\frac{C}{(\sin \psi)^{1+a}} \frac{I_1(y^*)}{(y^*)^2} = 1, \label{eq:yStarConstraint}
\end{align}
with 
\begin{align}
	C \equiv \frac{1}{6\sqrt{3}} \lp \frac{\omega_{p0} }{\Omega_0} \rp \lp \frac{\omega_p L }{c} \rp \frac{1}{\chi_e^5}.
\end{align}
Here, the first quantity in parentheses is the plasma overdensity parameter (assumed less than 1 for a tenuous plasma), and the second is the ratio of the system size to the plasma skin depth.

\subsubsection{Asymptotically-valid fit for $I_1$}

To facilitate solving for $y^*$ in Eq.~(\ref{eq:yStarConstraint}), it helps to have an asymptotically valid approximation scheme for $I_1(y)$.
At high $y$, and thus typically high $x$, we have from Eq.~(\ref{eq:K53IntegralApproximationHighX}):
\begin{align}
	\bar{\eta}(x)  \approx \sqrt{\frac{\pi}{2}} x^{1/2} e^{-x} \text{ for } x \gg 1. 
\end{align}
Thus, at high $y$, we have:
\begin{align}
	I_1 %&\rightarrow  \int_{0}^\infty dw \, w^2 \, e^{-w} \sqrt{\frac{\pi}{2}} \left(\frac{y}{w^2}\right)^{1/2} e^{-y/w^2}\\
	&\rightarrow  \sqrt{\frac{2}{3}} \pi y e^{-3 y^{1/3}/2^{2/3}} \text{ as } y\rightarrow \infty.
\end{align}

At low $y$, and thus low $x$, we have:
\begin{align}
	\bar{\eta}(x) \approx 2^{2/3} \Gamma\left(\frac{2}{3}\right) x^{1/3} \text{ for } x \ll 1. \label{eq:K53IntegralApproximationLowX}
\end{align}
Substitution into $I_1$ then yields:
\begin{align}
	I_1 %&\rightarrow  \int_{0}^\infty dw\,  w^2 e^{-w} 2^{2/3} \Gamma\left(\frac{2}{3}\right) \left(\frac{y}{w^2}\right)^{1/3}\\
	&\rightarrow 2^{2/3} \Gamma\left(\frac{2}{3}\right) \Gamma\left(\frac{7}{3}\right)  y^{1/3} \text{ as } y\rightarrow 0.
\end{align}

\begin{figure*}
	\centering
	\includegraphics[width=\linewidth]{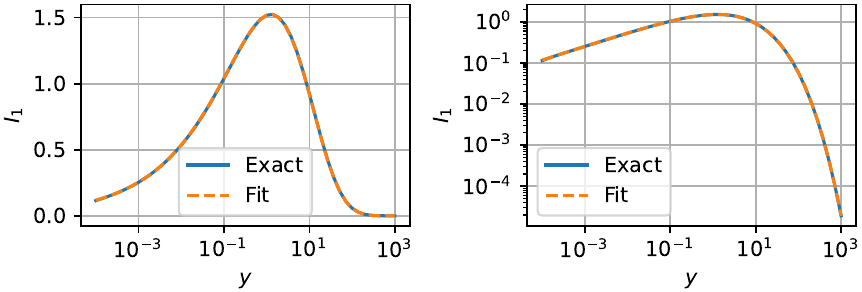}
	\caption{The integral quantity $I_1(y)$ that is required to calculate the radiation absorption in the ultrarelativistic limit, as a function of normalized frequency $y$: exact result (Eqs.~\ref{eq:I1Exact}-\ref{eq:xFromy}) and fit (Eq.~\ref{eq:I1Fit}).
		The right plot shows the same result on a log-log plot, demonstrating asymptotic agreement for large $y$.}
	\label{fig:I1Fit}.
\end{figure*}

Combining these limits and adding a fit parameter, as in the case of $\lambda$, yields an asymptotically-valid fit function:
\begin{align}
	I_1 &\approx y^{1/3} e^{-3 y^{1/3}/2^{2/3}} \left[2.56 + 4.92 y^{1/3}+ \sqrt{\frac{2}{3}} \pi y^{2/3}  \right]. \label{eq:I1Fit}
\end{align}
This fit is compared with the numerically calculated value in Fig.~\ref{fig:I1Fit}.

\subsubsection{Fit for $y^*$}

\begin{figure}
	\centering
	\includegraphics[width=\linewidth]{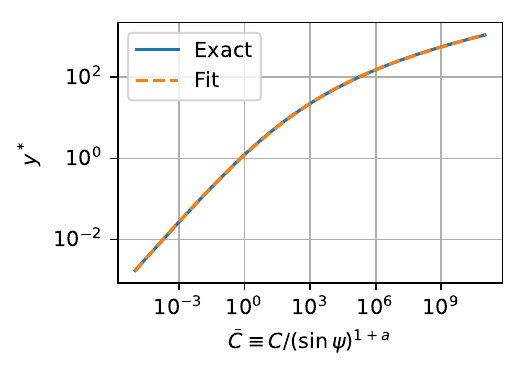}
	\caption{(Solid) Numerical solution for $y^*$ as a function of $\bar{C} \equiv C / (\sin \psi)^{1+a}$, from Eq.~(\ref{eq:yStarConstraint}) given Eq.~(\ref{eq:I1Fit}).
	(Dotted) Fit for $y^*$ from Eq.~(\ref{eq:yStarFit}).}
	\label{fig:yStarFromC}.
\end{figure}

Using this approximate value of $I_1$, it is very fast to numerically solve Eq.~(\ref{eq:yStarConstraint}) for $y^*$.
The result, which depends only on the combined parameter $\bar{C} \equiv C / (\sin \psi)^{1+a}$, is shown in Fig.~\ref{fig:yStarFromC}.
We can see that $y^*$ follows approximate power-law scaling relations with $\bar{C}$ in the limit of large and small $\bar{C}$, with a transition region in between.
Thus, $y^*$ fits well to the function:
\begin{align}
	y^* \approx 1.21 \bar{C}^{0.613 -0.387 \tfrac{\bar{C}^{0.113}}{2.26 + C^{0.113}}}. \label{eq:yStarFit}
\end{align}

\subsection{Summary of ultrarelativistic synchrotron-induced diffusion model}

It is useful at this point to summarize the complete model for diffusion from synchrotron emission in a highly relativistic plasma.
The diffusion model for a mixed-opacity plasma in coordinates of relativistically normalized momentum $u$ and pitch angle $\psi$ consists of Eqs.~(\ref{eq:FundamentalDiffusion}), (\ref{eq:RadiationFrequencyAndMetric}), and (\ref{eq:GammaUltraRelativistic}-\ref{eq:DUltraRelativistic}).
However, these can be reduced to a one-dimensional diffusion equation in $w \equiv u / \chi_\text{bb}$ for each value of $\psi$:
\begin{align}
	\pa{f}{t} &= \frac{1}{w^2}\pa{}{w} \cdot \left[\bar{\nu}_{R0} w^4  \sin^2 \psi \left(  f + (1-\lambda) \pa{f}{w} \right) \right] \label{eq:DiffusionW},
\end{align}
where
\begin{align}
	\bar{\nu}_{R0} = \chi_\text{bb} \nu_{R0}.
\end{align}

This diffusion equation depends on the opacity function $\lambda$, representing the fraction of radiation emitted at $(w,\psi)$ that escapes the plasma.
It turns out that $\lambda$ is a function only of $x^* = y^*(\psi,C)/w^2$.
The function $\lambda(x^*)$ is defined by Eq.~(\ref{eq:LambdaUltraRelativistic}) and well approximated by Eq.~(\ref{eq:UltrarelativisticLambdaFit}).
In turn, $y^*(\psi,C)$ is a function of the pitch angle $\psi$ and opacity parameter $C$ only through the combination $\bar{C} \equiv C / (\sin \psi)^{1+a}$.
The function $y^*(\bar{C})$ is defined as the solution to Eq.~(\ref{eq:yStarConstraint}), which is well-approximated by Eq.~(\ref{eq:yStarFit}).

\section{Behavior of the Diffusion} \label{sec:Heuristics}

The steady state of the plasma depends on the balance of diffusion and drag, which in turn depends on the function $\lambda$.
At $\lambda = 1$, all the radiation escapes the plasma, and so there is no diffusion, only drag, and the distribution collapses toward the origin.
The $u_\perp$-dependence of the drag term causes the distribution to elongate along $u_\parallel$ and form a population inversion in $u_\perp$, as described in \cite{Bilbao2023RadiationReaction,Bilbao2024RingMomentum,Zhdankin2023SynchrotronFirehose}.
In the opposite limit, at $\lambda = 0$, radiation is perfectly confined, and diffusion and drag balance to produce a Maxwell-J\"uttner distribution at the shared temperature of the blackbody radiation and electrons.

Crucially, however, $\lambda$ is not constant for all electrons, but depends strongly on the momentum.
From Fig.~\ref{fig:UltrarelativisticLambda}, we see that $\lambda$ rapidly transitions from 1 to 0 around $x^* \sim 1$.
Furthermore, from Eq.~(\ref{eq:xFromy}), $x^* = y^* / w^2$, where $w \equiv u/\chi_e$ is the ratio of the particle energy to the thermal energy.
As a result, the radiation escape fraction has an extremely strong dependence on the electron energy.

We can thus identify three main regimes. 
In the optically thin limit of $y^* \lesssim 1$, which occurs for $C \lesssim 1$, drag largely dominates the dynamics, which reduce to the results of \cite{Bilbao2023RadiationReaction,Bilbao2024RingMomentum,Zhdankin2023SynchrotronFirehose}: namely, the development of an unstable ring distribution with a population inversion along $w_\perp$ and a firehose-unstable $(P_\parallel > P_\perp)$ pressure tensor. 
Conversely, in the optically thick limit of $y^* \gtrsim 10^2$, corresponding to $C \gtrsim 10^5$, diffusion and drag balance for many thermal $e$-foldings, and the distribution remains largely thermal.

In the third, optically mixed regime of $1 \lesssim y^* \lesssim 10^2$, i.e. $1 \lesssim C \lesssim 10^5$, the balance of diffusion and drag is energy-dependent.
For such plasmas, the distribution will experience primarily thermalization to the blackbody temperature for $w \lesssim \sqrt{y^*}$, and primarily drag for $w \gtrsim \sqrt{y^*}$.
Since $y^*$ depends weakly on $\psi$, this transition point is anisotropic, occurring at lower values of $w$ for higher values of $\psi$, i.e. as $w_\perp$ increases relative to $w_\parallel$.
This anisotropy in the opacity, combined with the one-dimensional nature of the diffusion, turns out to have important consequences for the long-time behavior of the solution.

\subsection{Approximate Quasi-Steady State} 
For a fixed value of $C$, we can explore the long time behavior by looking at the steady-state solution to Eq.~(\ref{eq:DiffusionW}).
Such a solution not a true steady state of the radiating plasma, since the opacity properties of the plasma really depend on the plasma distribution itself.
The solution also neglects pitch-angle scattering, which occurs on a much slower timescale.
Nevertheless, it is useful as it represents a quasi-steady state of the particle distribution given fixed radiation properties of the plasma, on the timescale of the radiation-induced diffusion.

For a given value of the pitch angle, the quasi-steady-state distribution is approximately Maxwell-J\"uttner for $w < \sqrt{y^*}$, and rapidly suppressed for $w > \sqrt{y^*}$.
Thus, to understand the long-time stability properties, we can model the distribution as a truncated Maxwell-J\"uttner distribution.
However, critically, the 1D nature of the diffusion along $\hat{w}$ means that the electron pitch angle $\psi$ never changes.
Thus, the distribution must be separately normalized along each pitch angle $\psi$, i.e. for an initially isotropic distribution:
\begin{align}
	\int_0^\infty w^2 f(w,\psi) dw = \frac{1}{4\pi}.
\end{align}
This means that $f$ will take the general form:
\begin{align}
	f = \frac{N(\psi)}{8 \pi} e^{-w} H(\sqrt{y^*(\psi)} - w), \label{eq:TruncModel}
\end{align}
where 
\begin{align}
	N(\psi) &= \frac{\int_0^\infty w^2 e^{-w} dw}{\int_0^{\sqrt{y^*}} w^2 e^{-w} dw} = \frac{2}{2 - e^{-\sqrt{y^*}}(2 + 2\sqrt{y^*} + y^*)}.
\end{align}
With this very rough distribution, we can get an approximate idea of the stability features of the quasi-steady state.

\subsection{Ring Distribution}

A kinetic instability can often form if $\paf{f}{u_\perp} > 0$ for some point in the distribution; in fact, this is the basis for many fusion-system microinstabilities, including fast-ion instabilities \citep{Cook2017StimulatedEmission} and loss-cone instabilities in magnetic mirrors \citep{Kolmes2024LossconeStabilization}.
In terms of the $(u,\psi)$, this condition becomes: 
\begin{align}
	\pa{f}{w_\perp} &= \pa{w}{w_\perp} \pa{f}{w} + \pa{\psi}{w_\perp} \pa{y*}{\psi} \pa{f}{y^*}.
\end{align}

For a given value of $C$, Eq.~(\ref{eq:yStarFit}) shows that $y^*$ follows an approximate power law given by:
\begin{align}
	y^* &\approx 1.2 \left(\frac{C}{(\sin \psi)^{a+1}}\right)^b,
\end{align}
where $0.226 < b < 0.613$.
Thus, plugging in for $f$ from the truncated Maxwell-J\"uttner model, we have:
\begin{align}
	\pa{f}{w_\perp} &= \sin \psi \left[-1 + \frac{1}{w} \frac{\cos^2\psi}{\sin^2 \psi}  A(y^*(\psi)) \right] f,
\end{align}
where:
\begin{align}
	A(y^*) &\equiv 0.6 (a+1) b \frac{(y^*)^{3/2}}{2 e^{\sqrt{y^*}} - (2 + 2 \sqrt{y^*} + y^*)}.
\end{align}
We can thus see that at low $\psi$ and $w$, there is a tendency towards population inversion.
As $C$ gets bigger, $y^*$ gets bigger, $A(y^*)$ gets smaller, and this inversion is pushed to lower values of $w$ and $\psi$.

\subsection{Pressure Tensor} \label{sec:TruncPressure}

The truncated Maxwellian can also be used to calculate the degree of pressure anisotropy in steady state.
The nondimensional pressure tensor components for a kinetic distribution of particles in the relativistic limit are given by:
\begin{align}
	P_{\perp} &\equiv \int d\ve{w} \frac{w^x w^x}{|w|} f(\ve{w})\\
	&= \int (w^2 \sin \theta d\psi d\phi dw) w \sin^2 \psi \cos^2 \phi f(w,\psi)\\
	&= \pi \int_0^{\pi} d\theta \sin^3 \psi \int_0^\infty w^3 f(w,\psi). \label{eq:Pp}
\end{align}
for the perpendicular pressure, and:
\begin{align}
	P_{\parallel} &\equiv \int d\ve{w} \frac{w^z w^z}{|w|} f(\ve{w})\\
	&= \int (w^2 \sin \theta d\psi d\phi dw) w \cos^2 \psi f(w,\psi)\\
	&= 2 \pi \int_0^{\pi} d\psi \sin \psi \cos^2 \psi \int_0^\infty w^3 f(w,\psi) \label{eq:Pl}
\end{align}
for the parallel pressure.
Note that for the Maxwell-J\"uttner distribution, $P_\perp = P_\parallel = 1$.

\begin{figure*}
	\centering
	\includegraphics[width=0.9\linewidth]{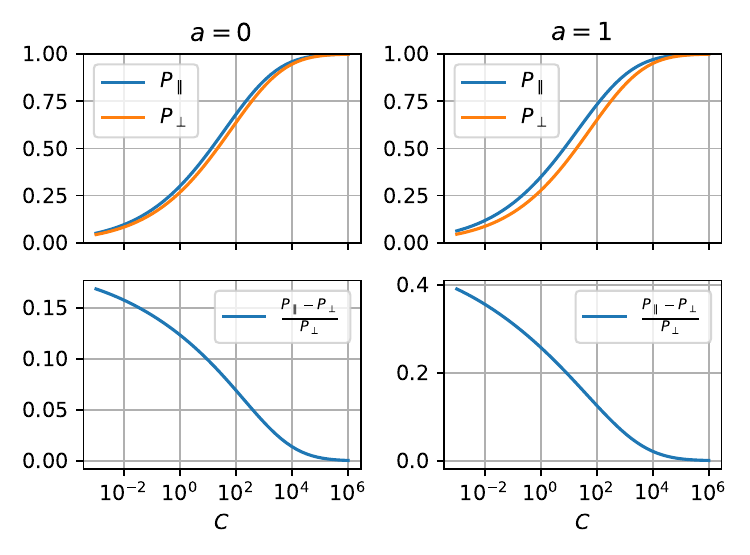}
	\caption{Pressure tensor components and pressure anisotropy as a function of $C$ for the approximate quasi-steady state electron distribution (the truncated Maxwell-J\"uttner model from Eq.~(\ref{eq:TruncModel})).
	Shown are the geometric cases of $a = 0$ (sphere) and $a=1$ (cylinder / slab).
	There is less anisotropy in a spherical plasma, and generally less anisotropy with increasing $C$.}
	\label{fig:PressureTensorTruncDist}.
\end{figure*}

For the truncated Maxwell-J\"uttner distribution model from Eq.~(\ref{eq:TruncModel}), we can calculate:
\begin{align}
	\int_0^w w^3 f(w,\psi) &= \frac{1}{4\pi}\frac{6 - e^{-\sqrt{y^*}}(6 + 6\sqrt{y^*} + 3y^* + (y^*)^{3/2})}{2 - e^{-\sqrt{y^*}}(2 + 2\sqrt{y^*} + y^*)}. \label{eq:fw3Integral}
\end{align}
Further analytic progress is not particularly informative; however, plugging Eq.~(\ref{eq:yStarFit}) into Eq.~(\ref{eq:fw3Integral}) produces an expression that can be easily numerically integrated over $\psi$ to determine the pressure anisotropy as a function of $C$.
The result is shown in Fig.~\ref{fig:PressureTensorTruncDist}, both for $a=0$ and $a=1$.
At lower $C$ the distribution is characterized by lower quasi-steady-state pressure, with a higher degree of firehose-like pressure anisotropy ($P_\parallel / P_\perp > 1$).
The anisotropy is more pronounced for the cylindrical / slab case of $a=1$, than the sphere case of $a=0$, as the radiation emitted by high-$u_\parallel$ electrons is less well trapped in the latter case.

\section{Numerical Simulations} \label{sec:Simulations}

\begin{figure*}
	\centering
	\includegraphics[width=\linewidth]{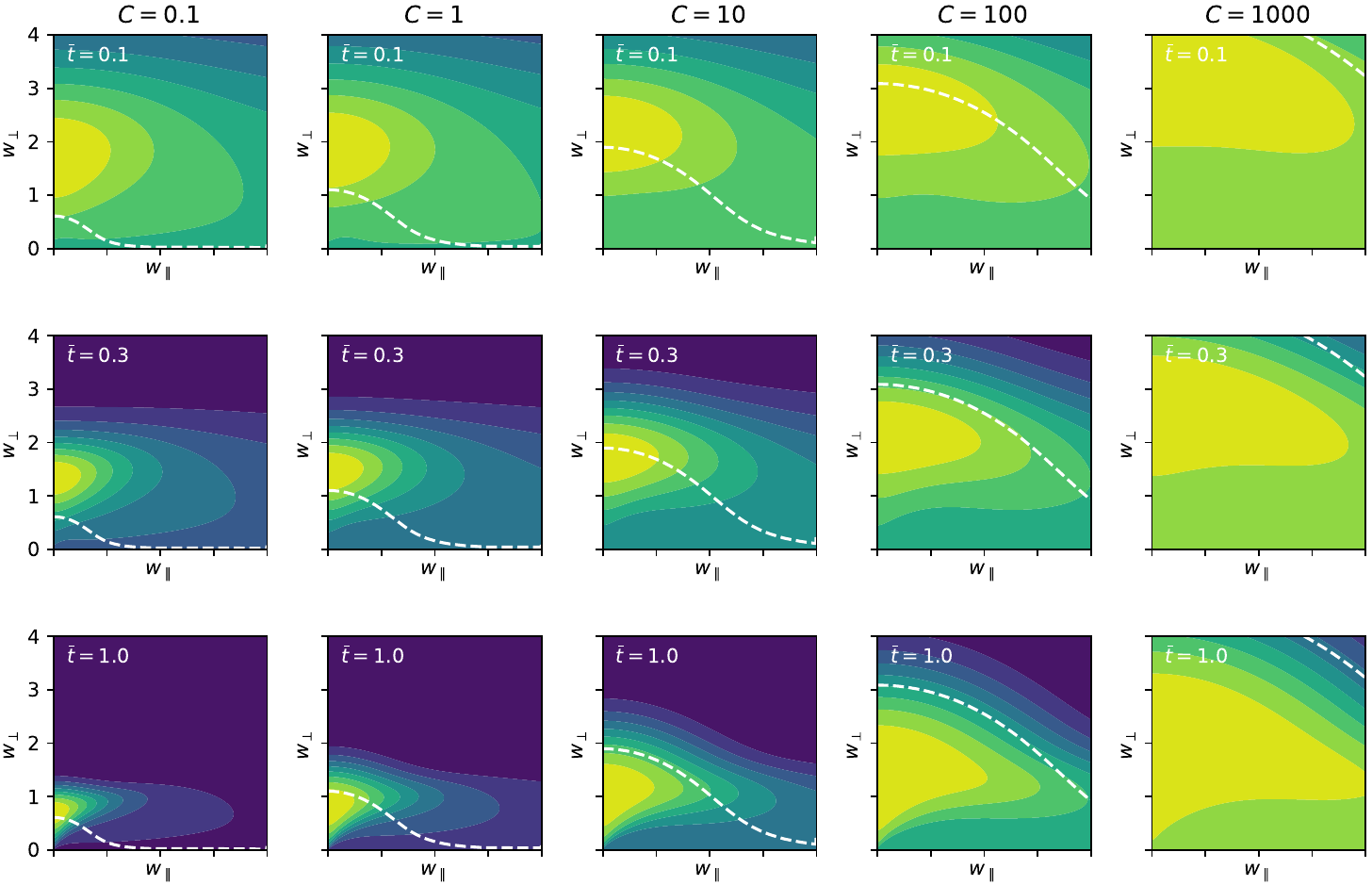}
	\caption{Evolution of an initial Maxell-J\"uttner distribution in the presence of synchrotron emission and absorption for several values of the opacity parameter $C$.
		Plotted is the density scaled relative to a Maxwell-J\"uttner distribution, i.e. $f(w) e^w$, where $w \equiv u / \chi_\text{bb}$.
		Each distribution is shown at several multiples of the collision time $\bar{t} \equiv \bar{\nu}_{R0} t$.
		The normalization of each plot is arbitrary and only reflects areas of relative density in the distribution.
		The dashed line represents the point at which $x^* = 1$, i.e. where $\lambda = 1/2$.
		Several trends are clear from the simulations.
		First, diffusion and drag occur much faster out on the tail, so that the quasi-steady state ``propagates inward'' from large $w$.
		Second, the tendency toward population inversion in $w_\perp$ is seen across all populations, but is notably weaker as $C$ increases and the plasma grows more opaque.
		Third, the distribution is elongated along $w_\parallel$ for all values of $C$.
		And fourth, as steady-state is approached, the $x^* = 1$ line provides an approximate divider between a low-energy Maxwell-J\"uttner distribution, and tail suppression at high energies, justifying the crude truncated-tail model used to examine the plasma stability properties.}
	\label{fig:AstroSims}.
\end{figure*}

Having established expectations using a crude analytical model for the solution, we now turn to direct numerical simulations of the Fokker-Planck equation.
In \cite{Ochs2024ElectronTail}, Eq.~(\ref{eq:FundamentalDiffusion}) was simulated directly in two dimensions.
However, that study considered the mildly relativistic regime, where pitch angle scattering was significant.
Because pitch angle scattering is negligible for very relativistic plasmas, the diffusion operator is effectively one-dimensional, and two-dimensional simulations are subject to numerical instabilities.

Instead, to observe the evolution of the electron distribution under the influence of emission and absorption of radiation, we simulate one-dimensional diffusion for each value of the pitch angle $\psi$.
In the coordinate $w \equiv u/\chi_\text{bb}$, this takes the form of Eq.~(\ref{eq:DiffusionW}), with zero-flux boundary conditions at both $w\rightarrow 0$ and $w \rightarrow \infty$.
Such a simulation scheme was constructed using the DOLFINx \citep{Scroggs2022ConstructionArbitrary} finite-element library, using a backward-Euler time-advance.

As in \cite{Bilbao2023RadiationReaction,Bilbao2024RingMomentum,Zhdankin2023SynchrotronFirehose}, we considered the evolution of an initial Maxwell-J\"uttner distribution, which was taken to be at the radiation blackbody temperature.
To get accurate results for the pressure tensor, simulations were performed on a grid from $w = 0$ to $w = 13$, as it takes many $e$-foldings for $w^3 f_\text{MJ}(w)$ to become small.

Results of the simulations are shown in Fig.~\ref{fig:AstroSims} for $a = 0$ and several values of the plasma opacity parameter $C$.
In the figure, time runs from top to bottom, plotted in terms of the normalized time $\bar{t} \equiv \bar{\nu}_{R0} t$.
Also shown is the line of $x^* = 1$, the rough boundary of the low-energy ``opaque'' region and the high-energy ``transparent'' region of phase space.

The simulations reveal several facts.
First, diffusion and drag occur much faster out on the tail; this means that that the quasi-steady state appears to ``propagate inward'' from large $w$.
Second, while the tendency toward population inversion in $w_\perp$ (as found in \cite{Bilbao2023RadiationReaction,Bilbao2024RingMomentum}) is seen across all populations, it becomes notably weaker as $C$ increases and the plasma grows more opaque.
Third, as found in \cite{Zhdankin2023SynchrotronFirehose}), the distribution is elongated along $w_\parallel$.
And finally, at long times, the $x^* = 1$ line provides an approximate divider between a low-energy Maxwell-J\"uttner distribution, and tail suppression at high energies, thus justifying the crude truncated-tail model used in Sec.~\ref{sec:Heuristics}. 

\begin{figure*}
	\centering
	\includegraphics[width=\linewidth]{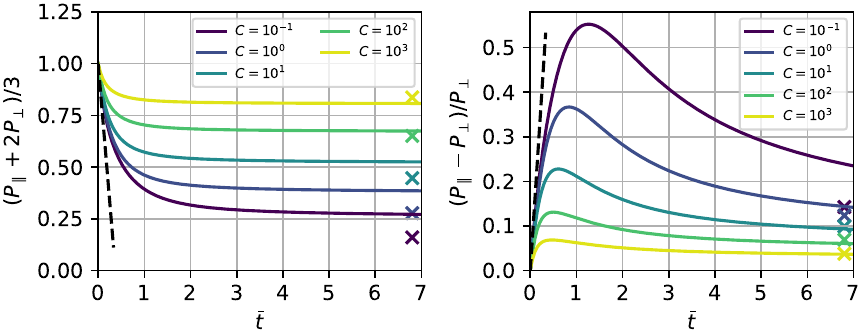}
	\caption{Average pressure $(P_\parallel + 2P_\perp)/3$ and pressure anisotropy $(P_\parallel - P_\perp)/P_\perp$ as a function of time for the simulations in Fig.~\ref{fig:AstroSims}.
	The black dashed line shows the optically-thin, short-time result from Ref.~\cite{Zhdankin2023SynchrotronFirehose}, i.e. Eqs.~(\ref{eq:ZhdankinPAv}-\ref{eq:ZhdankinDeltaP}).
	The ``$\times$'' marks represent the quasi-steady-state predictions of the truncated Maxwellian model in Sec.~\ref{sec:TruncPressure}.
	As the opacity parameter $C$ increases, the pressure plateaus earlier and at a higher value.
	Anisotropy also increases with increasing $C$.
	Interestingly, the pressure anisotropy first increases and then decreases to its quasi-steady-state value.
	The truncated Maxwellian model does a fairly good job at predicting the pressure anisotropy until $C \lesssim 10$.}
	\label{fig:PressureTensorSims}.
\end{figure*}

The simulations can also be used to calculate the pressure tensor components from Eqs.~(\ref{eq:Pp}) and (\ref{eq:Pl}), using builtin DOLFINx integration features to perform the inner integral (over $w$), and scipy's builtin Simpson integration method for the outer integral (over $\psi$).
It is most instructive to consider the average pressure $(P_\parallel + 2 P_\perp)/3$, and the pressure anisotropy $(P_\parallel - P_\perp)/P_\perp$
The results are shown in Fig.~\ref{fig:PressureTensorSims}.
At short times, the results closely follow those found in \cite{Zhdankin2023SynchrotronFirehose}, namely (noting that \cite{Zhdankin2023SynchrotronFirehose}'s $\tau_\text{cool} = 1/(3 \bar{\nu}_{R0})$):
\begin{align}
	\frac{P_\parallel + 2 P_\perp}{3} &= 1 - \frac{8}{3} \bar{t} \label{eq:ZhdankinPAv}\\
	\frac{P_\parallel - P_\perp}{P_\perp} &= 1 + \frac{8}{5} \bar{t}. \label{eq:ZhdankinDeltaP}
\end{align}
However, as $C$ increases, the components of $\ve{P}$ plateau at an increasingly early time, and thus at a higher level.
Interestingly, while the average pressure monotonically decreases before plateauing, the pressure anisotropy initially increases rapidly, and then decreases again toward a steady-state plateau value, exhibiting an intermediate-time maximum at $\bar{t} \equiv \bar{\nu}_{R0} t \lesssim 1$.
We see that the plateau values for the average pressure and pressure anisotropy agree fairly well with those predicted by the truncated Maxwell-J\"uttner model until $C \lesssim 10$.

\section{Conclusions}

In this paper, we have found how electron distributions evolve in relativistic plasmas with some degree of opacity, and determined the parameters that control the (frequency-dependent) opacity.
We have shown that some degree of ring distribution formation and pressure anisotropy persist even as the plasma becomes optically thick, but that these features are weakened as the critical opacity parameter $C$ becomes larger.
These opacity effects have the potential to stabilize distributions that would otherwise be driven unstable \citep{Bilbao2023RadiationReaction,Bilbao2024RingMomentum,Zhdankin2023SynchrotronFirehose} in optically thin plasmas.
Furthermore, the evolution of the electron distribution in such arbitrary-opacity plasmas could be relevant for magnetic reconnection processes as well, since radiative cooling is increasingly realized to be important for reconnection in a variety of astrophysical systems, including pulsar magnetospheres, pulsar wind nebulae and magnetar magnetospheres \citep{Uzdensky2011MagneticReconnection,Datta2024SimulationsRadiatively}, and is thus a very active area of research for reconnection in pulsed-power experiments \citep{Datta2024SimulationsRadiatively,Datta2024RadiativelyCooled}.
If there is asymmetry in the electron distribution, then emission and absorption of synchrotron radiation could also lead to current generation and magnetogenesis, as proposed for synchrotron-reflection current drive in tokamaks \citep{Dawson1982CurrentMaintenance} and inverse bremsstrahlung and synchrotron current drive in astrophysical discs \citep{Munirov2017InverseBremsstrahlung,Munirov2017RadiativeTransfer}, for which kinetic effects are of great importance \citep{fisch1980creating,fisch1987theory}.

Of course the present model has many simplifications, which may need to be addressed for more quantitative modeling of these systems.
First and foremost, the model assumed the plasma was homogeneous, so that plasma emission and absorption occurred with equal frequency throughout the plasma.
In a realistic plasma, photons emitted from the plasma edge would be more likely to escape than those emitted from the core, making the plasma effectively optically thinner at the edge than the center.
This would lead to a cooling of the edge relative to the core, as well as net radiation transport outward from the core to the edge, all of which is neglected in the current spatially uniform model.

Second, we have neglected the evolution of the plasma opacity, i.e. the fact that the opacity itself is dependent on the electron distribution.
In a realistic plasma, as the plasma grows colder, it allows a broader range of frequencies to escape, growing optically thinner, and thus cooling the plasma more.
The present work shows how one could self-consistently approach such a problem, but the details have yet to be worked out.
This consideration, combined with the neglect of the (much slower) pitch angle scattering near $u_\perp\sim 1$, means that great care has to be taken in applying the quasi-steady state solution.

Third, we have here assumed that the magnetic field has a single uniform value and orientation.
In a turbulent, magnetoactive plasma, there would more likely be a spectrum of local magnetic field strengths with somewhat randomized orientations.
This could dramatically change some of the results; for instance, while the local slowing would still drive anisotropy, the plasma opacity itself might cease to be anisotropic, reducing or eliminating the quasi-steady-state anisotropy of the electron distribution function.
This could quickly eliminate anisotropies in higher-opacity plasmas.

Finally, it is important to mention that even though we have been working in the relativistic limit, we have treated radiation drag as a continuous process, neglecting the increasing quantization of the photon emission at ultrarelativistic energies.
Thus, as discussed in \cite{Bilbao2024RingMomentum}, the present analysis should be understood to apply to relativistic plasmas $\gamma \gg 1$ where the photon emission energy $\hbar \omega \ll \gamma m c^2$.
If the latter condition is not satisfied, than an additional diffusion term appears related to the large-jump nature of the quantized photon emission.

Thus, while there are many ways the model could be made more detailed, the present paper points toward the rich variety of kinetic effects that can be expected in moderate-opacity, synchrotron-emitting relativistic plasmas.

\section*{Acknowledgments}

The author would like to thank Mike Mlodik and Nat Fisch for illuminating discussions.
This work was supported by DOE Grant No. DE-SC0016072.

%\begin{align}
%	\Dv &= \chi_0 \bvec D_\parallel (1+\cos 2 \theta) + D_\perp (1-\cos 2 \theta) & \frac{(D_\parallel - D_\perp) \sin 2  \theta}{x}  \\
%	\frac{(D_\parallel - D_\perp) \sin 2  \theta}{x}  & \frac{D_\perp (1+\cos 2 \theta) + D_\parallel (1-\cos 2 \theta)}{x^2} \evec
%\end{align}

\bibliographystyle{jpp}

%\bibliography{../../../Reading/allRefsZot.bib}

\begin{thebibliography}{19}
\expandafter\ifx\csname natexlab\endcsname\relax\def\natexlab#1{#1}\fi
\def\au#1{#1} \def\ed#1{#1} \def\yr#1{#1}\def\at#1{#1}\def\jt#1{\textit{#1}}
  \def\bt#1{#1}\def\bvol#1{\textbf{#1}} \def\vol#1{#1} \def\pg#1{#1}
  \def\publ#1{#1}\def\arxiv#1{#1}\def\org#1{#1}\def\st#1{\textit{#1}}

\bibitem[Bekefi(1966)]{Bekefi1966RadiationProcesses}
{\sc \au{Bekefi, George}} \yr{1966} {\em Radiation {{Processes}} in
  {{Plasmas}}\/}.  \publ{New York: {John Wiley and Sons}}.

\bibitem[Bilbao {\em et~al.\/}(2024)Bilbao, Ewart, Assun{\c c}ao, Silva \&
  Silva]{Bilbao2024RingMomentum}
{\sc \au{Bilbao, P.~J.}, \au{Ewart, R.~J.}, \au{Assun{\c c}ao, F.}, \au{Silva,
  T.} \& \au{Silva, L.~O.}} \yr{2024}  \at{Ring momentum distributions as a
  general feature of {{Vlasov}} dynamics in the synchrotron dominated regime}.
  \jt{Physics of Plasmas}  \bvol{31}~(5),  \pg{052112}.

\bibitem[Bilbao \& Silva(2023)]{Bilbao2023RadiationReaction}
{\sc \au{Bilbao, P.~J.} \& \au{Silva, L.~O.}} \yr{2023}  \at{Radiation
  {{Reaction Cooling}} as a {{Source}} of {{Anisotropic Momentum
  Distributions}} with {{Inverted Populations}}}.  \jt{Physical Review Letters}
   \bvol{130}~(16),  \pg{165101}.

\bibitem[Cook {\em et~al.\/}(2017)Cook, Dendy \&
  Chapman]{Cook2017StimulatedEmission}
{\sc \au{Cook, J. W.~S.}, \au{Dendy, R.~O.} \& \au{Chapman, S.~C.}} \yr{2017}
  \at{Stimulated {{Emission}} of {{Fast Alfv{\'e}n Waves}} within
  {{Magnetically Confined Fusion Plasmas}}}.  \jt{Physical Review Letters}
  \bvol{118}~(18),  \pg{185001}.

\bibitem[Datta {\em et~al.\/}(2024{\natexlab{{\em a\/}}})Datta, Chandler,
  Myers, Chittenden, Crilly, Aragon, Ampleford, Banasek, Edens, Fox, Hansen,
  Harding, Jennings, Ji, Kuranz, Lebedev, Looker, Patel, Porwitzky, Shipley,
  Uzdensky, {Yager-Elorriaga} \& Hare]{Datta2024RadiativelyCooled}
{\sc \au{Datta, R.}, \au{Chandler, K.}, \au{Myers, C.~E.}, \au{Chittenden,
  J.~P.}, \au{Crilly, A.~J.}, \au{Aragon, C.}, \au{Ampleford, D.~J.},
  \au{Banasek, J.~T.}, \au{Edens, A.}, \au{Fox, W.~R.}, \au{Hansen, S.~B.},
  \au{Harding, E.~C.}, \au{Jennings, C.~A.}, \au{Ji, H.}, \au{Kuranz, C.~C.},
  \au{Lebedev, S.~V.}, \au{Looker, Q.}, \au{Patel, S.~G.}, \au{Porwitzky, A.},
  \au{Shipley, G.~A.}, \au{Uzdensky, D.~A.}, \au{{Yager-Elorriaga}, D.~A.} \&
  \au{Hare, J.~D.}} \yr{2024{\natexlab{{\em a\/}}}}  \at{Radiatively cooled
  magnetic reconnection experiments driven by pulsed power}.  \jt{Physics of
  Plasmas}  \bvol{31}~(5),  \pg{052110}.

\bibitem[Datta {\em et~al.\/}(2024{\natexlab{{\em b\/}}})Datta, Crilly,
  Chittenden, Chowdhry, Chandler, Chaturvedi, Myers, Fox, Hansen, Jennings, Ji,
  Kuranz, Lebedev, Uzdensky \& Hare]{Datta2024SimulationsRadiatively}
{\sc \au{Datta, Rishabh}, \au{Crilly, Aidan}, \au{Chittenden, Jeremy~P.},
  \au{Chowdhry, Simran}, \au{Chandler, Katherine}, \au{Chaturvedi, Nikita},
  \au{Myers, Clayton~E.}, \au{Fox, William~R.}, \au{Hansen, Stephanie~B.},
  \au{Jennings, Chris~A.}, \au{Ji, Hantao}, \au{Kuranz, Carolyn~C.},
  \au{Lebedev, Sergey~V.}, \au{Uzdensky, Dmitri~A.} \& \au{Hare, Jack~D.}}
  \yr{2024{\natexlab{{\em b\/}}}}  \at{Simulations of radiatively cooled
  magnetic reconnection driven by pulsed power}.  \jt{Journal of Plasma
  Physics}  \bvol{90}~(2),  \pg{905900215}.

\bibitem[Dawson \& Kaw(1982)]{Dawson1982CurrentMaintenance}
{\sc \au{Dawson, John~M.} \& \au{Kaw, Predhiman~K.}} \yr{1982}  \at{Current
  {{Maintenance}} in {{Tokamaks}} by {{Use}} of {{Synchrotron Radiation}}}.
  \jt{Physical Review Letters}  \bvol{48}~(25),  \pg{1730--1733}.

\bibitem[Fisch(1987)]{fisch1987theory}
{\sc \au{Fisch, Nathaniel~J.}} \yr{1987}  \at{Theory of current drive in
  plasmas}.  \jt{Reviews of Modern Physics}  \bvol{59}~(1),  \pg{175--234}.

\bibitem[Fisch \& Boozer(1980)]{fisch1980creating}
{\sc \au{Fisch, N.~J.} \& \au{Boozer, A.~H.}} \yr{1980}  \at{Creating an
  {{Asymmetric Plasma Resistivity}} with {{Waves}}}.  \jt{Physical Review
  Letters}  \bvol{45}~(9),  \pg{720--722}.

\bibitem[Kolmes {\em et~al.\/}(2024)Kolmes, Ochs \&
  Fisch]{Kolmes2024LossconeStabilization}
{\sc \au{Kolmes, E.J.}, \au{Ochs, I.E.} \& \au{Fisch, N.J.}} \yr{2024}
  \at{Loss-cone stabilization in rotating mirrors: Thresholds and
  thermodynamics}.  \jt{Journal of Plasma Physics}  \bvol{90}~(2),
  \pg{905900203}.

\bibitem[Kolmes {\em et~al.\/}(2020)Kolmes, Ochs, Mlodik \&
  Fisch]{Kolmes2020MaxEntropy}
{\sc \au{Kolmes, E.J.}, \au{Ochs, I.E.}, \au{Mlodik, M.E.} \& \au{Fisch, N.J.}}
  \yr{2020}  \at{Maximum-entropy states for magnetized ion transport}.
  \jt{Physics Letters A}  \bvol{384}~(13),  \pg{126262}.

\bibitem[Melrose(1980)]{Melrose1980PlasmaAstrophysics}
{\sc \au{Melrose, Donald~B.}} \yr{1980} {\em Plasma Astrophysics. 1: {{The}}
  Emission, Absorption and Transfer of Waves in Plasmas\/}, ,  \vol{vol.~1}.
  \publ{New York, NY: {Gordon and Breach}}.

\bibitem[Mlodik {\em et~al.\/}(2023)Mlodik, Munirov, Rubin \&
  Fisch]{Mlodik2023SensitivitySynchrotron}
{\sc \au{Mlodik, M.~E.}, \au{Munirov, V.~R.}, \au{Rubin, T.} \& \au{Fisch,
  N.~J.}} \yr{2023}  \at{Sensitivity of synchrotron radiation to the
  superthermal electron population in mildly relativistic plasma}.  \jt{Physics
  of Plasmas}  \bvol{30}~(4),  \pg{043301}.

\bibitem[Munirov \& Fisch(2017{\natexlab{{\em
  a\/}}})]{Munirov2017InverseBremsstrahlung}
{\sc \au{Munirov, Vadim~R.} \& \au{Fisch, Nathaniel~J.}}
  \yr{2017{\natexlab{{\em a\/}}}}  \at{Inverse {{Bremsstrahlung}} current
  drive}.  \jt{Physical Review E}  \bvol{96}~(5),  \pg{053211}.

\bibitem[Munirov \& Fisch(2017{\natexlab{{\em
  b\/}}})]{Munirov2017RadiativeTransfer}
{\sc \au{Munirov, Vadim~R.} \& \au{Fisch, Nathaniel~J.}}
  \yr{2017{\natexlab{{\em b\/}}}}  \at{Radiative transfer dynamo effect}.
  \jt{Physical Review E}  \bvol{95}~(1),  \pg{013205}.

\bibitem[Ochs {\em et~al.\/}(2024)Ochs, Mlodik \& Fisch]{Ochs2024ElectronTail}
{\sc \au{Ochs, Ian~E.}, \au{Mlodik, Mikhail~E.} \& \au{Fisch, Nathaniel~J.}}
  \yr{2024} Electron {{Tail Suppression}} and {{Effective Collisionality}} due
  to {{Synchrotron Emission}} and {{Absorption}} in {{Mildly Relativistic
  Plasmas}},  \arxiv{arXiv: 2407.09716}.

\bibitem[Scroggs {\em et~al.\/}(2022)Scroggs, Dokken, Richardson \&
  Wells]{Scroggs2022ConstructionArbitrary}
{\sc \au{Scroggs, Matthew~W.}, \au{Dokken, J{\o}rgen~S.}, \au{Richardson,
  Chris~N.} \& \au{Wells, Garth~N.}} \yr{2022}  \at{Construction of arbitrary
  order finite element degree-of-freedom maps on polygonal and polyhedral cell
  meshes}.  \jt{ACM Trans. Math. Softw.}  \bvol{48}~(2).

\bibitem[Uzdensky(2011)]{Uzdensky2011MagneticReconnection}
{\sc \au{Uzdensky, Dmitri~A.}} \yr{2011}  \at{Magnetic {{Reconnection}} in
  {{Extreme Astrophysical Environments}}}.  \jt{Space Science Reviews}
  \bvol{160}~(1-4),  \pg{45--71}.

\bibitem[Zhdankin {\em et~al.\/}(2023)Zhdankin, Kunz \&
  Uzdensky]{Zhdankin2023SynchrotronFirehose}
{\sc \au{Zhdankin, Vladimir}, \au{Kunz, Matthew~W.} \& \au{Uzdensky,
  Dmitri~A.}} \yr{2023}  \at{Synchrotron {{Firehose Instability}}}.  \jt{The
  Astrophysical Journal}  \bvol{944}~(1),  \pg{24}.

\end{thebibliography}

\clearpage\newpage

\end{document}